# Preliminaries on the Accurate Estimation of the Hurst Exponent Using Time Series

Ginno Millán, *IEEE Member*, Román Osorio-Comparán, and Gastón Lefranc, *IEEE Senior Member*

*Abstract*—This article explores the required amount of time series points from a high-speed computer network to accurately estimate the Hurst exponent. The methodology consists in designing an experiment using estimators that are applied to time series addresses resulting from the capture of high-speed network traffic, followed by addressing the minimum amount of point required to obtain in accurate estimates of the Hurst exponent. The methodology addresses the exhaustive analysis of the Hurst exponent considering bias behaviour, standard deviation, and Mean Squared Error using fractional Gaussian noise signals with stationary increases. Our results show that the Whittle estimator successfully estimates the Hurst exponent in series with few points. Based on the results obtained, a minimum length for the time series is empirically proposed. Finally, to validate the results, the methodology is applied to real traffic captures in a high-speed computer network.

## I. Introduction

FRACTAL processes are indicative of stochastic behavior that is invariant to changes in dimensional scales and temporary [1]–[5]. These processes are applied as models in various fields of science [6]–[10]. In the area of computer networks and telecommunications systems, they are used to model traffic on LAN, MAN, WAN, WWW, and different technologies of cellular and wireless networks [11]–[15].

In all these studies, traffic is measured and then analyzed to determine whether or not it fits a fractal behavior. The traffic traces used in all analyzes are made up of hundreds of thousands of samples and often with long capture times. For offline traffic studies [16] these lengths and waiting times are acceptable, however for real-time network administration applications with QoS metrics based on the precise estimation of *H* [17] these lengths and capture times are too large.

This article first studies the behavior of the estimators applied to short-term time series and then addresses the problem of the minimum length required to obtain accurate estimates of *H*. Therefore, the aim is to obtain high precision with a minimum length, as opposed to [18], where the author raises the impossibility of determining a minimum length for fractal time series without losing their intrinsic properties.

G. Millán is with the Facultad de Ingeniería y Tecnología, Universidad San Sebastián, Puerto Montt, Chile (e-mail: ginno.millan@uss.cl).
R. Osorio-Comparán is with the IIMAS, Universidad Nacional Autónoma de México (e-mail: roman@unam.mx).
G. Lefranc is with the Escuela de Ingeniería Eléctrica, Pontificia Universidad Católica de Valparaíso, Chile (e-mail: gaston.lefranc@pucv.cl).

To develop the problem posed by the estimation of *H*, the accuracy must be comparable with the problems of long series; where the accuracy is based on metrics such as bias (*b*), standard deviation ($\sigma$), and the Mean Squared Error (MSE).

The article then addresses the following problem: Given the specification(s) of *b* or MSE, what should be the minimum length, $N_{min}$, of the time series that satisfies them?; that is, suppose that the stochastic process *X* has a Hurst exponent *H*, the minimum length $N_{min}$ of the time series must be found so that each proposed $N_{min}$ presents the estimated Hurst exponents $H_e$ that is similar to the Hurst *H* exponent of the original process *X*.

## II. Theoretical Framework

### A. Fractal Processes

A process is fractal if its distribution of probabilities is invariant to temporal dilation and compression of its amplitude.

Let $Y = \{Y_t\}_{t \in I}$, where $I = \mathbb{R}$ or $\mathbb{R}^+$, is a real-value stochastic process. It is said that *Y* is a fractal process if and only if, there is $H \in \mathbb{R}$, such that for all $a \in \mathbb{R}^+$, the following relationship $\{Y_{at}\}_{t \in I} =_d \{a^H Y_t\}_{t \in I}$ is fulfilled, where $=_d$ means equality in their probability distributions [19].

Generally the interest is focused on fractal processes with stationary type increases with $H > 0$. This definition is known as strict.

A second definition is obtained by invariance in second-order statistics.

Let $Y_t$ be a continuous-time stochastic process. $Y_t$ is said to be a second-order self-similar process, if and only if it complies with $E(Y_t) = a^{-H}E(Y_{at})$, for all $a > 0$, $t \geq 0$ and $0 < H < 1$, where $E(\cdot)$ is the median of the process, $Var[Y_t] = a^{-2H}Var[Y_{at}]$, for all $a > 0$, $t \geq 0$ and $0 < H < 1$, where $Var[\cdot]$ is the variance of stochastic process *Y*, and its autocorrelation function, $R_{zz}(\cdot)$, behaves according to the relationship $R_{zz}(t, s) = a^{-2H}R_{zz}(at, as)$ for all $a > 0$, $t \geq 0$, and $0 < H < 1$ [19].

Computer networks require a discrete version of the definition of fractal processes.

The model is defined as $X = \{X_t\}_{t \in \mathbb{Z}}$ which a discrete process due to the sampling of a continuous random signal.

*X* is strictly self-similar, if and only if, $X =_d m^{1-H}\Gamma_m(X)$, with $0 < H < 1$, for all $m \in \mathbb{N}$, where $\Gamma(\cdot)$ represents the block aggregation process that receives a time series of length *N* as input and provides an output as a time series of length *N/m* [20].

The second version of this definition is that of second-order self-similarity in the exact sense.

Formally, $X$ is an exact second-order self-similar process with $\sigma^2$ variance of the process, if its autocovariance function $\rho(k)$ has the following form for the range $0.5 < H < 1$ [21]

$$\rho(k) = \frac{1}{2}\sigma^2[(k+1)^{2H} - 2k^{2H} + (k-1)^{2H}], \quad \text{for all } k \geq 1. \quad (1)$$

A stochastic process with an autocovariance function given by (1) also satisfies the following restrictions

$$\text{Var}[X] = m^{2-2H}\text{Var}[\Gamma(X)], \quad (2)$$

$$\text{Cov}[\Gamma_m(X_t), \Gamma_m(X_{t+k})] = m^{2-2H}\text{Cov}[X_t, X_{t+k}], \quad (3)$$

for its variance and covariance, respectively.

In the field of computer networks, a relaxed version of (1) is used. A stochastic process $X$ is asymptotically self-similar to the second order if the correlation factor $\Gamma_m(X)$ when $m \to \infty$ is equal to the self-similar stochastic process of discrete-time, that is to say (1).

In particular, let $R_{zz}(k) = \rho(k)/\alpha^2$ denote the autocorrelation function. For $0 < H < 1$, $H \neq 0.5$, it holds [21]

$$R_{zz}(k) \sim H(2H-1)k^{2H-2}, \quad \text{when } k \to \infty. \quad (4)$$

In particular, if $0.5 < H < 1$, $R_{zz}(k)$ asymptotically behaves as $ck^{-\beta}$ for $0 < \beta < 1$, where $c > 0$ is a constant, $\beta = 2 - 2H$, and we have [21]

$$\sum_{k=-\infty}^{\infty} R_{zz}(k) = \infty. \quad (5)$$

That is, the autocorrelation function decays slowly (that is, hyperbolically) which is the essential property that causes it to not be summable.

When $R_{zz}(k)$ decays hyperbolically so that condition (5) holds, we call the corresponding stationary process $X$, Long-range dependence (LRD) processes [21].

### B. On the Hurst Exponent Estimation

Different methods have been proposed to estimate $H$; these can be classified into methods developed for time domain, frequency domain, and time-scale methods.

Among the time domain methods is the R/S statistic, the aggregated variance method, absolute value method, variance of the residuals, the Higuchi´s method, the Modified Variance of Allan (MAVAR), the scale window variation, the Whittle estimator, etc. [22].

The periodogram method and the modified periodogram method [17], the Geweke and Porter-Hudak method [18], are in the frequency domain class which takes advantage of the characteristic power-law behavior of the self-similar processes in the neighborhood of their origin [22].

Finally, the time-scale methods include all wavelet-based estimators such as the Abry and Veitch´s method and their variants [22], [23].

### III. WORKING METHODOLOGY

The study of fractality by analyzing the value of $H$ allows its presence and its degree of persistence to be detected. The methodology developed for the calculation of the different estimators for different lengths of time series explained below.

To apply the estimators, the $N$ time series must be obtained. Synthetic signals with known $H$ are obtained through the simulation of a series of fractional Gaussian noise (fGn) [24] with stationary increases using the Davies and Harte method described in detail in [25].

For the experiments, 200 traffic traces with $H \in \{0.5, 0.6, 0.7, 0.8, 0.9\}$ and lengths $N = \{2^i, i = 6, 7, 8, 9, 10, 11, 12, 13, 14, 15, 16\}$ where considered that is 11000 fractal signals (200 * $H$ * $N$). For each sets of estimates of a particular $H$, the calculation of the statistics described above was performed

- Bias: $b = H_0 - \bar{X}$, where $H_0$ is the nominal value of $H$ and $\bar{X}$ is the average of the values of the $X$ process.
- Standard deviation $\sigma$.
- Mean Square Error $\text{MSE} = N^{-1}\sum_{i=1}^{N}(X_i - H_0)^2$.

Then, based on these three estimators, a minimum series length, $N_{\min}$, is proposed from the estimates that consider

- $b \sim 0.03$, and
- $\sigma \sim 0.01$.

Together with the above, estimates based on $b$ and $\sigma$ are classified as follows

- High precision: when $b \leq 0.03$ and $\sigma \leq 0.01$.
- Acceptable: when $0.03 < b < 0.05$ and $\sigma \leq 0.02$.
- Biased (but not unacceptable): when $b > 0.1$.

Once the minimum lengths for the fGn series is obtained, the results obtained are then applied to real traffic traces. For these series, designated as $Z$, of length $M$, such that $M \gg N_{\min}$ the procedure is

1. Let $t_0, \ldots, t_k$ be a sequence of points on the x-axis, where it is true that $t_{i+1} > t_i$ and $t_{i+1} - t_i < N_{\min}$, for each block of $Z$ of length $N_{\min}$, $\{Z_j\}_{j=t_i}^{t_i+N_{\min}-1}$, be the estimate of $H$ ($H_e$), that is, $H_{e,t_i}^{N_{\min}}(\cdot)$, until $t_k + N_{\min} > M$, for any $k$.

2. It is speculated that $N_{\min}$ is chosen correctly, if $t_i$ versus $H_{e,t_i}^{N_{\min}}(\cdot)$ is plotted, the result show a signal with a little variation, ie the variation should be equal to the value of $\sigma$.

3. $H_{e,t_i}^{N_{\min}}(\cdot)$ is applied over a joint set of the series, i.e. $\bigcup_{i=1}^{Nm^{-1}}\{\Psi_i^m\}$, to obtain a sufficient amount of $H$ estimators; amount described by the set $\{H_{e,1}^m, \ldots, H_{e,1}^{jm}\}$. The term $\Psi_i^m$ is justified as follows. The correct length $N_{\min}$ is directly related to the convergence of a series, for this reason the study of this relationship is carried out as follows, but not before remembering that the convergence of an estimator is obtained by disaggregating the original $Z$ series in blocks of size $m \ll M$, with the objective of obtain a set $Z$ defined by $Z = \{\Psi_i^m\}$, where the set $\Psi_i^m$ is defined by $\Psi_i^m = \{Z_{im}\}$.

## IV. SIMULATION AND RESULTS

Fig. 1 shows $H_e$ for a fGn series using Whittle estimator, Fig. 2 shows $H_e$ using the Abry and Veitch´s method, Fig. 3 $H_e$ using periodogram method, and Fig. 4 $H_e$ using the R/S statistic, in all cases the software used is SELFIS [26], [27].

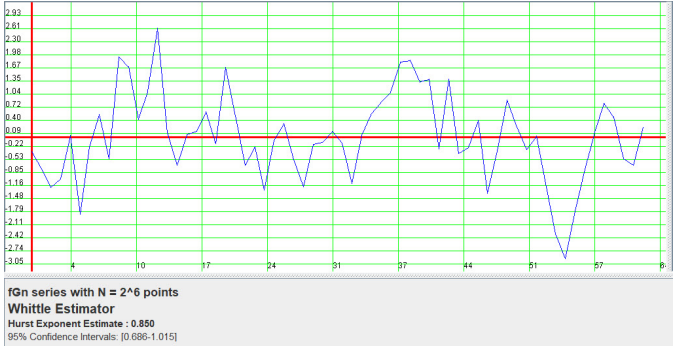

(a)

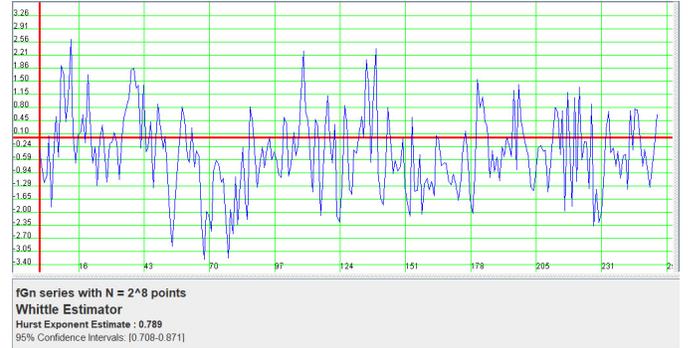

(b)

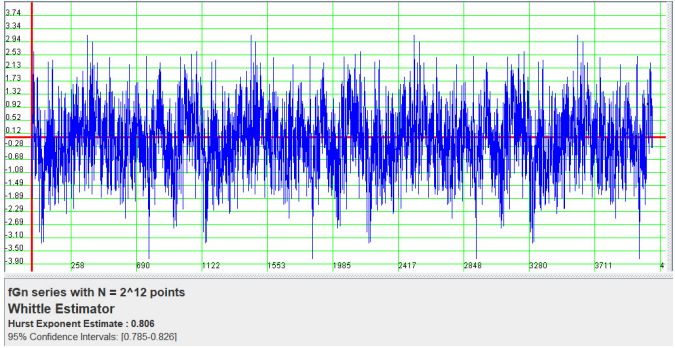

(c)

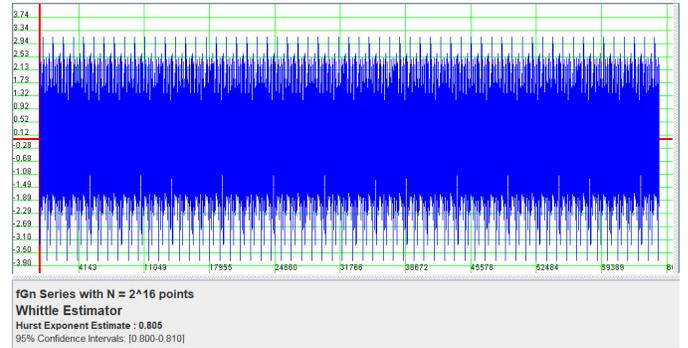

(d)

Fig. 1. Synthetic fGn series with $N = 2^6$ (a), $N = 2^8$ (b), $N = 2^{12}$ (c), and $N = 2^{16}$ (d) points, respectively, using the Whittle estimator.

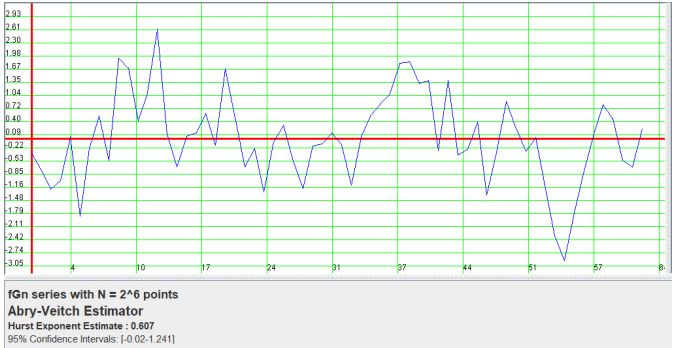

(a)

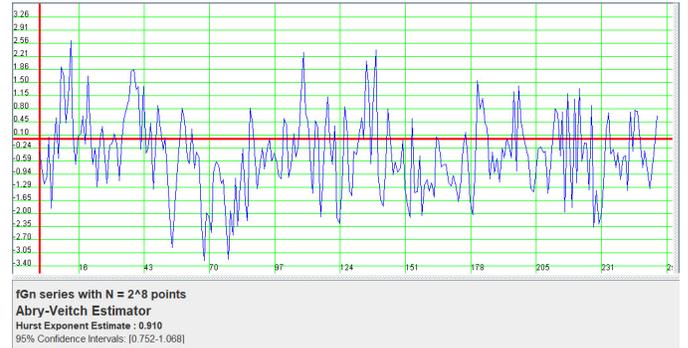

(b)

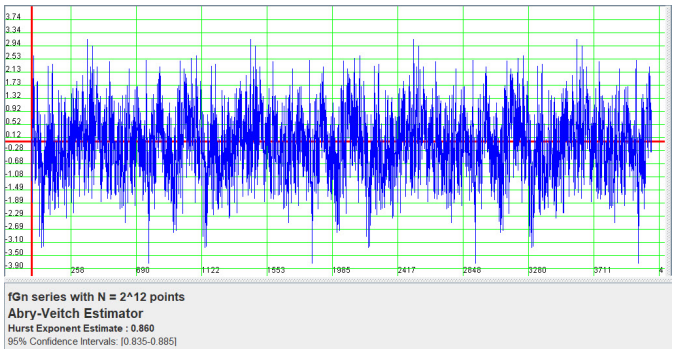

(c)

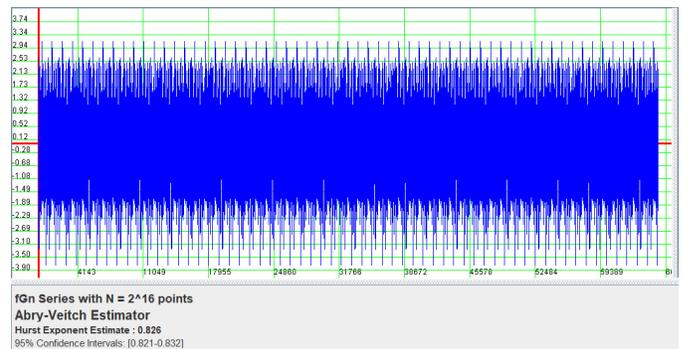

(d)

Fig. 2. Synthetic fGn series with $N = 2^6$ (a), $N = 2^8$ (b), $N = 2^{12}$ (c), and $N = 2^{16}$ (d) points, respectively, using the Abry and Veitch´s method.

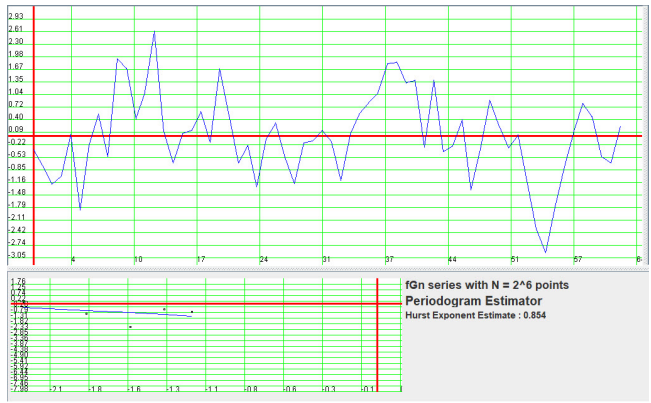
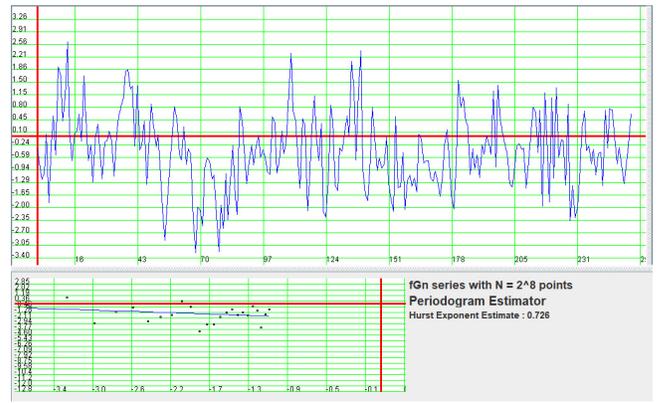

(a)                      (b)

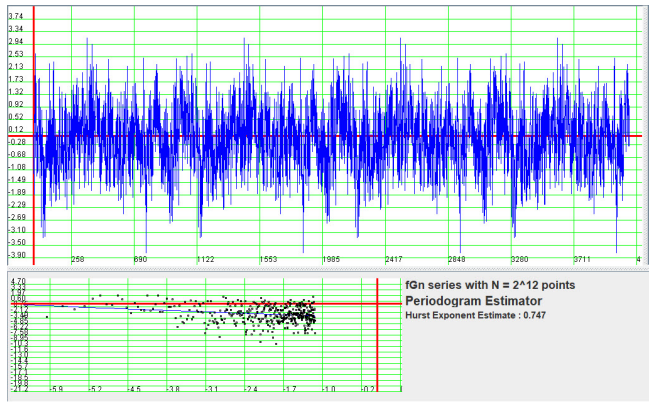
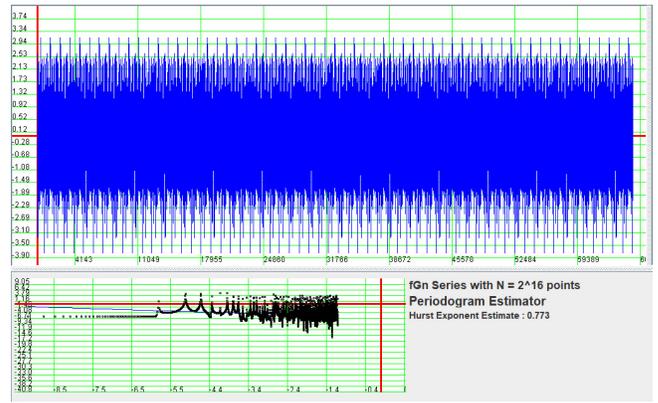

(c)                      (d)

Fig. 3. Synthetic fGn series with $N = 2^6$ (a), $N = 2^8$ (b), $N = 2^{12}$ (c), and $N = 2^{16}$ (d) points, respectively, using the periodogram method.

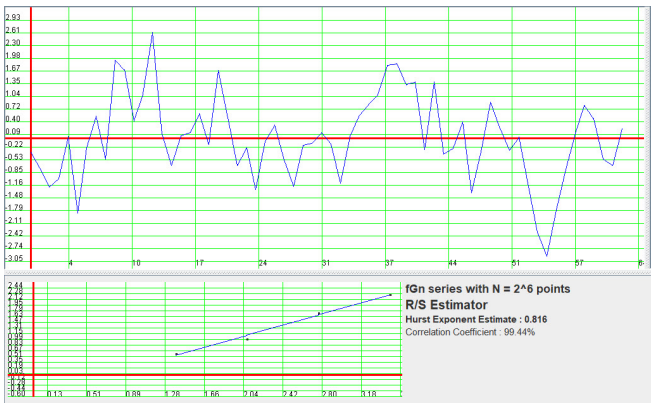
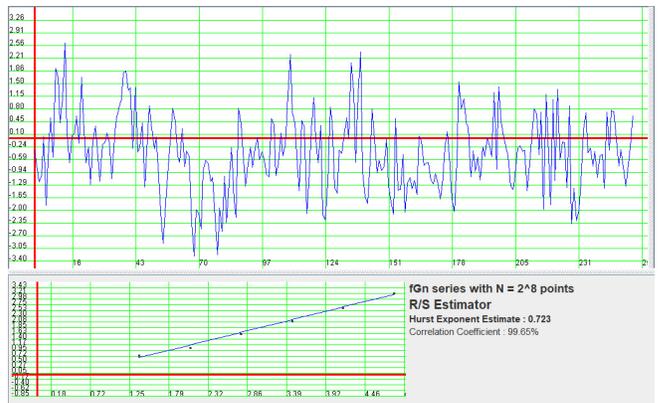

(a)                      (b)

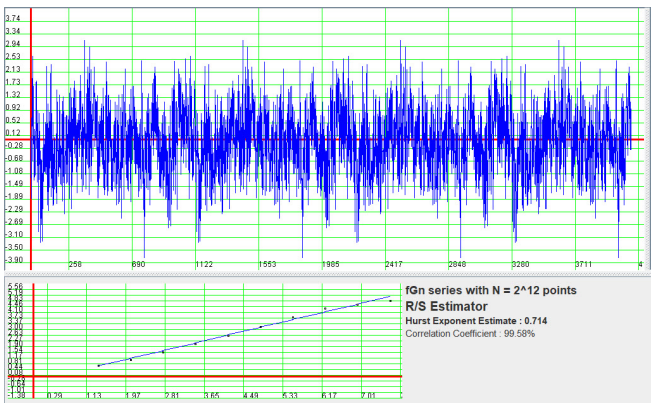
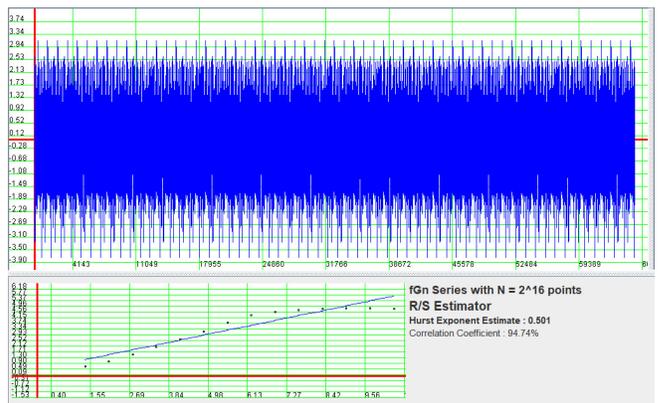

(c)                      (d)

Fig. 4. Synthetic fGn series with $N = 2^6$ (a), $N = 2^8$ (b), $N = 2^{12}$ (c), and $N = 2^{16}$ (d) points, respectively, using the R/S statistic.

Table I shows a summary of the obtained results.

TABLE I
SUMMARY OF THE RESULTS OBTAINED IN FIGURES 1, 2, 3, AND 4

| Used Technique | $H$ Value for the Synthetic Fractional Gaussian Noise (fGn) Series with | | | | Respective Figure |
|---|---|---|---|---|---|
| | $N = 2^6$ | $N = 2^8$ | $N = 2^{12}$ | $N = 2^{16}$ | |
| Whittle Estimator | 0.850 | 0.789 | 0.806 | 0.805 | 1(a), 1(b), 1(c), and 1(d) |
| Abry and Veitch´s Method | 0.607 | 0.910 | 0.806 | 0.826 | 2(a), 2(b), 2(c), and 2(d) |
| Periodogram Method* | 0.854 | 0.726 | 0.747 | 0.773 | 3(a), 3(b), 3(c), and 3(d) |
| R/S Statistic | 0.816 | 0.723 | 0.714 | 0.501 | 4(a), 4(b), 4(c), and 4(d) |

\* The periodogram method finds its based in the behavior of the near origin of the Spectral Power Density (PSD).

Based on the results obtained from the experiments for the interval $2^6 < N < 2^{16}$, the estimates of $H$ deliver acceptable results, i.e. $0.03 < b < 0.05$ and $\sigma \leq 0.02$ and for $N \geq 2^{13}$ the estimates are highly precise, i.e. $b \sim 0.03$ and $\sigma \sim 0.01$. Note that from Section III, that high precision is defined by $b \leq 0.03$ and $\sigma \leq 0.01$, a fact by which the previous approximations for $b$ and $\sigma$ do not introduce error.

The analysis of the results obtained shows that the estimates of exponent $H$ using the Whittle estimator show high precision and low variability for the standard length series reported in the literature, for $N \sim 2^{10}$ points. Note: most of the literature consulted uses $N = 2^{10}$ instead of $N \sim 2^{10}$.

Based on the analysis of the results obtained if $N < 2^{12}$ and $H \geq 0.8$, the estimates are accurate.

On the other hand, the results obtained using the Whittle´s estimator are compared with the results obtained using the Abry and Veitch´s method.

The results based on Whittle´s estimator are more accurate than their wavelet counterparts for short series in the context of the synthetic fGn series of the study.

Fig. 5 shows the bias for all the methods used on the fGn series for different data lengths. This allows us to observe the variation of the differences estimates of value $H$ delivered by each analysis method. Note from in Fig. 5 that the Whittle´s estimator and Abry and Veitch´s method have a better behavior for short length series, $N < 2^{10}$ points, than the other analysis techniques. Also, the Whittle estimator behaves with less irregularity than Abry and Veitch´s method for short length series ($N < 2^{10}$) and for series with lengths $N \geq 2^{10}$, the bias presented by both techniques is not significant.

For the Abry and Veitch´s method, $b$ behaves irregularly for the short series and stabilizes with $N \geq 2^{14}$.

For the Whittle estimator, the bias behaves irregularly for the short series and stabilizes with $N \geq 2^{11}$.

The other techniques (periodogram method and R/S statistic) exhibit a completely irregular behavior and a very high bias and unless $N \geq 2^{16}$ (see Fig. 5), which is why its estimates are not considered acceptable.

The bias of the R/S statistic does not show a behavior that stabilizes, while the periodogram method is stabilized for a high $N$, that is, $N \geq 2^{16}$ (see Fig. 5). Note that this graphic method has high variability regardless of the length of the time series.

With the same objective pursued by the study of the bias of the estimators considered in Fig. 5, Fig. 6 shows the behavior of the standard deviation of the estimators for the traces with variable $N$ length in the interval $2^6 < N < 2^{16}$.

Note from Fig. 6 that the Whittle estimator is the one that presents the least variability and for $N \leq 2^8$ the estimates are not precise enough to be considered. However, this does not mean that the method can be ruled out for time series of the few point ($N \leq 2^8$). This should be part of a large discussion.

Abry and Veitch´s method follow in precision, and the length required for the series can be considered as relatively identical when considering variability.

Finally, when considering both bias (Fig. 5) and standard deviation (Fig. 6), the best estimator for short time series is the Whittle estimator, with high precision results for $N \geq 2^8$, the Abry and Veitch´s method presents good precision for $N \sim 2^{12}$, the periodogram method has acceptable estimates for $N > 2^{15}$ and the R/S statistic has very biased estimates for $2^6 < N < 2^{16}$.

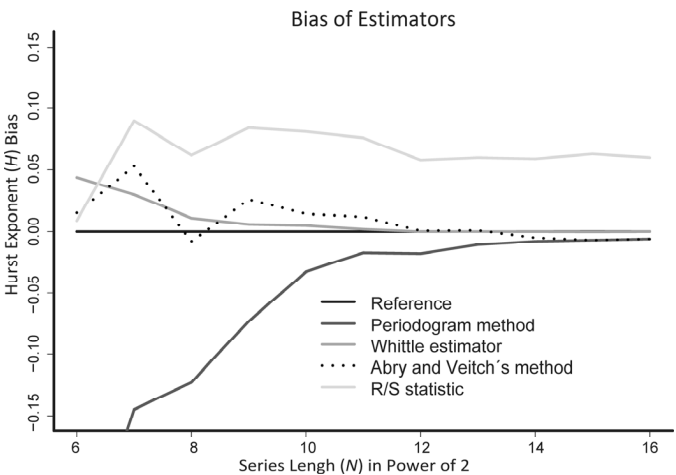

Fig. 5. Bias behavior for all analysis techniques considered.

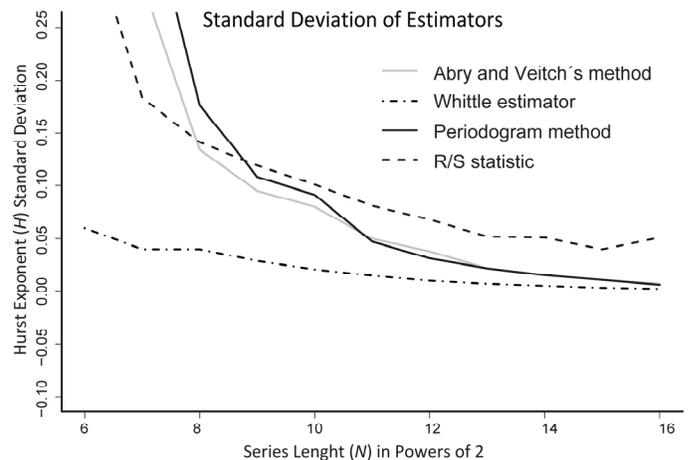

Fig. 6. Standard deviation for all analysis techniques considered.

## V. APPLICATION TO REAL TRAFFIC TRACES

### A. Generalities and Specifications of the Test Scenario

Based on the previous results on synthetic fGn traces, partial conclusions were applied to real traffic traces.

The traces were obtained from the core switch from the Departamento de Ingeniería Eléctrica of the Universidad de Santiago de Chile with the Center of Operations Control of the University corporate network.

For the capture of the traces, Wireshark [28] is used and they are transformed into time series using MATLAB, to construct ordered pairs (time of arrival at the sniffer, size of the captured frame). Capture considers the storage and processing capacity of 24 traffic hours of the Departamento de Ingeniería Eléctrica of the Universidad de Santiago de Chile considering bidirectional traffic, which translates into a total of $N = 2^{32}$ points using the following procedure

1. Let $t_0, \ldots, t_k$ be a sequence of points on the x-axis, where it is true that $t_{i+1} > t_i$ and $t_{i+1} - t_i < N_{min}$ is verified. Unless this time $N_{min} = 2^8$ points is defined, which implies starting with a time series given by $t_{i+1} > t_i$ and $t_{i+1} - t_i = 2^8$.
2. Each aggregate block in the time series it is estimated in reverse, the blocks change to a shorter length to approximate total real traffic flows.

Considerations about the experiment considering regarding point 1 and 2 above

- The value $N_{min} = 2^8$ results from a general appreciation of the results obtained from the combination and interpretation of the results obtained in Sections II and III; therefore, it is not a value chosen for convenience of analysis.
- Applying a disaggregation process can suggest, as a process contradictory to the theory, but this research seeks a $N_{min}$ that for both theoretical and practical results leads to a useful tool in real-time. However, an aggregation procedure cannot be homologated to a real-time one for the estimation of $H$ and the decision making for network administration and their implications and consequences.

### B. Obtained Results

Table II presents the general results for the $H$ estimation ($H_e$) techniques considered

TABLE II
GENERAL OVERVIEW OF ESTIMATES OF $H$ FOR REAL TRAFFIC TRACES

| Used Technique | Estimate Value of $H$ ($H_e$) | Correlation Coefficient | Confidence Interval |
|---|---|---|---|
| Whittle Estimator | 0.806 | Does not apply | 95% [0.785 – 0.826] |
| Abry and Veitch´s Method | 0.860 | Does not apply | 95% [0.835 – 0.885] |
| Periodogram Method | 0.947 | Does not apply | Does not apply |
| R/S Statistic | 0.714 | 99.58% | Does not apply |

From Table II, it is observed that both the Whittle estimator and that Abry and Veitch´s technique are consistent with the estimates of $H$ reported in [22]. However, for the Abry and Veitch´s method a high non-accused variability is observed in said reference while the periodogram method overestimates the value of $H$ and the R/S statistic exhibits irregular behavior.

Finally, the main results obtained can be grouped as follows considering short, medium and long lengths of the series, i.e. $N < 2^{10}$, $2^{10} \leq N < 2^{24}$, and $2^{24} \leq N \leq 2^{32}$ points of the time series under analysis, respectively.

Detail of main results obtained

1. The Whittle´s estimator behaves in a good way when it comes to short length time series that exhibit both minimal bias and variability.
2. The Abry and Veitch´s method behaves acceptably when is applied to series of medium length.
3. The periodogram method behaves acceptably if it is applied to time series of medium length.
4. The R/S statistic shows a high bias and therefore it is not suitable for application to short length time series.
5. Based on analysis performed, the minimum length for the Whittle estimator is $N \geq 2^{10}$.
6. Based on analysis performed, the minimum length for the Abry and Veitch´s method is $N \geq 2^{13}$.
7. On the basis of performed analyzes the minimum length for the periodogram method is $N \geq 2^{15}$.
8. It is not possible to estimate a minimum series length for the R/S statistic due to its high bias and its variability when it comes to length of $N \geq 2^{16}$. This case is particularly interesting due to its widespread use (only as a graphic technique). In this regard, considering the number of previous required, its use in estimation can be considered good but has a great computational cost.
9. It is not possible to find a length for the time series that respond to all the estimation techniques of $H$ considered in this research.

## VI. DISCUSSION OF RESULTS

### A. Overview

Traditional process-based traffic models with short-range dependency do not provide details on the behavior of flows in current high-speed data networks. Consequently, it is necessary to rethink the study of computer networks charging models that consider fractal entry traffics: since their requirements impose new challenges to network engineering, especially in buffering strategies of active equipment and estimation of yields.

This research presents preliminaries of the behavior of the most used estimators in the literature in synthetic time series using fGn series and then extrapolates results to real traffic traces obtained from a high-speed LAN network based on the IEEE 802.3ab standard.

Based on the behavioral study of the time series used of both fGn and real IEEE 802.3ab by applying bias analysis, standard deviation behavior, and Hurst exponent estimation, an attempt was made to determine a minimum length called $N_{\min}$ to establish a broad criterion of fractal analysis for time series involved.

Regarding the above, establishing a broad criterion for fractal analysis has an extra problem associated with it, namely, since it involves analysis based on successive approximations (iterations) by the algorithms that represent each of the considered estimators a mean convergence analysis of the behavior of the estimators is necessary. This fact is clearly manifested with the results shown in the final list of the previous section, since, although the sizes of short, medium and long series are established, certain results, such as those shown in points 6, 7 and 8, make it clear that such work will not be possible without convergence analysis. In this regard, this analysis will endow the investigation with an essential element: the speed with which each succession converges to its limit. Therefore, this concept is, from a practical point of view, completely necessary since we are working with sequences of successive approximations of an iterative method for each $H$ exponent analysis methodology. Furthermore, this analysis can make the difference between needing thousand or a million iterations for each algorithm committed in each estimator of $H$ exponent.

Research as such does not result in a specified length, since each type of analysis responds, even with the same objective, to different scenarios in which it is a question of checking the fractal nature of traffic flows in high-speed computer networks of different technologies.

It is necessary to emphasize that given the foregoing, the Research as such does not result in a length as specified, since each type of analysis responds, even with the same objective, to different scenarios in which it is a question of checking the fractal nature of traffic flows in high-speed computer networks of different technologies.

*B. Specific Results*

The results listed at the end of Section V are reviewed and interpreted again but considering the above

1. The Whittle´s estimator behaves in a good way when it comes to short length time series that exhibit both minimal bias and variability.
2. The Abry and Veitch´s method behaves acceptably when is applied to series of medium length.
3. The periodogram method behaves acceptably if it is applied to time series of medium length.
4. The R/S statistic shows a high bias and is therefore not suitable for application to short length time series.
5. Based on analyzes performed, the minimum length for the Whittle estimator is $N \geq 2^{10}$.
6. On the basis of performed analyzes, the minimum length for the Abry and Veitch´s method is $N \geq 2^{13}$.
7. On the basis of performed analyzes the minimum length for the periodogram method is $N \geq 2^{15}$.
8. It is not possible to estimate a minimum series length for the case of the R/S statistic due to its high bias and its variability when it comes to length of $N \geq 2^{16}$. This case is particularly interesting due to its widespread use (only as a graphic technique). In this regard, considering the number of previous required, its use in estimation can be considered good but has a great computational cost.
9. It is not possible to find a length for the time series that respond to all the estimations techniques of $H$ considered in this research.

## VII. CONCLUSION AND FUTURE DIRECTIONS

In this research we presented an analysis of traffic flows in high-speed computer networks using a minimum quantity of time series points that must contain estimates of the Hurst exponent. An experiment using estimators applied to time series provides an accurate determination of the Hurst exponent. In the exhaustive analysis of the Hurst exponent, bias behavior, standard deviation, and Mean Squared Error using fGn noise signals with stationary increases were considered.

The behavior of most estimators used in the literature, in synthetic time series using fGn series followed by extrapolation, results in real traffic traces obtained from a high-speed network based on the IEEE 802.3ab standard. Based on this behavior by applying bias analysis, average convergence analysis and Hurst exponent estimation, an attempt is made to determine a minimum length called $N_{\min}$ and to establish a broad criterion of fractal analysis.

The results obtained with the Whittle estimator allowed the Hurst exponent to be obtained in series with a few points. Then, a minimum length for the series is empirically proposed for each estimator considered.

Finally, to validate the results, the methodology was applied to actual traffic captures in a high-speed network based on the IEEE 802.3ab standard. The Whittle's estimator behaves in a good way when it comes to short series and long series that exhibit both minimal bias and variability.

The following two future tasks are proposed to be carried out as soon possible

1. In methodology section (Section III) specifies how to obtain the $N_{min}$ length for the fGn series. The results obtained are applied to real traffic traces. Traffic traces are designated by $Z$ and have a length $M$ such that $M \gg N_{\min}$.
   Then, the analysis presented establishes the need to analyze the convergence of the estimators to strengthen our postulate that an optimal length is not feasible, but it is possible to speak of a minimum number of points for each series depending on the estimator chosen to analyze the Hurst exponent.
2. As shown in the simulation and results section (Section IV), it is absolutely necessary to look for a software tool other that presents with higher resolution the resolution of the estimators for the Hurst exponent for the time series that contain a quantity $N > 2^{12}$.